\documentclass[sigconf,natbib]{acmart}

\usepackage{bm}
\usepackage{booktabs}
\usepackage{url}
\usepackage{hyperref}
\usepackage{multirow}
\usepackage[ruled,vlined,boxed,linesnumbered]{algorithm2e}
\usepackage{caption}
\usepackage{graphicx}
\usepackage{subcaption}
\usepackage{enumitem}
\usepackage{balance}
\usepackage{caption}
\usepackage{float}

\AtBeginDocument{%
  }

\setcopyright{acmlicensed}
\copyrightyear{2018}
\acmYear{2018}
\acmDOI{XXXXXXX.XXXXXXX}
\acmConference[Woodstock '18]{Woodstock '18: ACM Symposium on Neural
  Gaze Detection}{June 03--05, 2018}{Woodstock, NY}
\acmISBN{978-1-4503-XXXX-X/2018/06}

\begin{document}


\title{EGA-V2: An End-to-end Generative Framework for Industrial Advertising} 


\author{Zuowu Zheng$^*$, Ze Wang$^*$, Fan Yang$^\dagger$, Jiangke Fan, Teng Zhang, Yongkang Wang, Xingxing Wang}

\affiliation{%
 \institution{Meituan, Shanghai, China}
 \city{}
 \country{}
}
\email{{zhengzuowu, wangze18, yangfan129, jiangke.fan, zhangteng09, wangyongkang03, wangxingxing04}@meituan.com}








\renewcommand{\shortauthors}{Zheng et al.}

\begin{abstract}
\renewcommand{\thefootnote}{\fnsymbol{footnote}}
\footnotetext[1]{Equal contribution.}
\renewcommand{\thefootnote}{\fnsymbol{footnote}}
\footnotetext[2]{Corresponding author.}
  \renewcommand{\thefootnote}{\arabic{footnote}}

Traditional online industrial advertising systems suffer from the limitations of multi-stage cascaded architectures, which often discard high-potential candidates prematurely and distribute decision logic across disconnected modules. While recent generative recommendation approaches provide end-to-end solutions, they fail to address critical advertising requirements of key components for real-world deployment, such as explicit bidding, creative selection, ad allocation, and payment computation.

To bridge this gap, we introduce End-to-End Generative Advertising (EGA-V2), the first unified framework that holistically models user interests, point-of-interest (POI) and creative generation, ad allocation, and payment optimization within a single generative model. Our approach employs hierarchical tokenization and multi-token prediction to jointly generate POI recommendations and ad creatives, while a permutation-aware reward model and token-level bidding strategy ensure alignment with both user experiences and advertiser objectives. Additionally, we decouple allocation from payment using a differentiable ex-post regret minimization mechanism, guaranteeing approximate incentive compatibility at the POI level. Through extensive offline evaluations we demonstrate that EGA-V2 significantly outperforms traditional cascaded systems in both performance and practicality. Our results highlight its potential as a pioneering fully generative advertising solution, paving the way for next-generation industrial ad systems.
\end{abstract}

\ccsdesc[500]{Information systems~Display advertising}

\keywords{Generative Advertising, Recommendation System, Preference Alignment}

\maketitle

\section{Introduction}
\begin{figure}[h]
	\centering
	\includegraphics[width=\linewidth,angle=0]{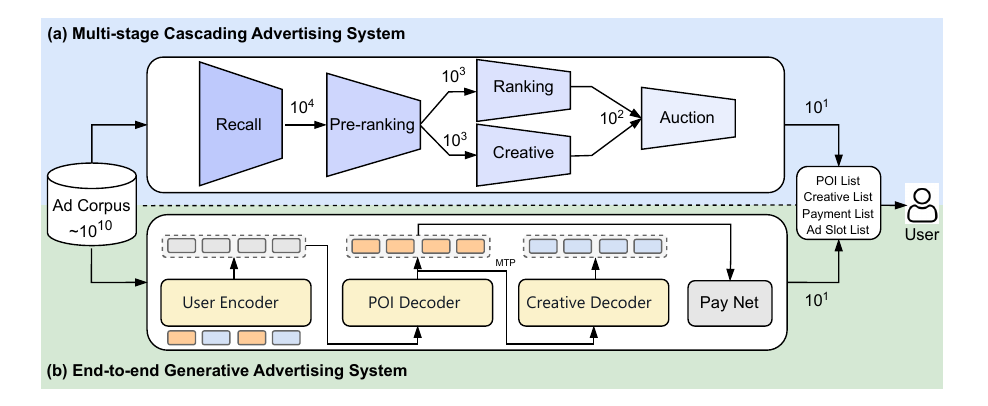}
	\caption{(a) A typical cascade advertising system. (b) Our proposed unified architecture for end-to-end generation.}
	\label{fig:pipline}
\end{figure}
Online advertising has become a critical revenue engine for major internet platforms, directly supporting the sustainability of a wide range of digital services. Each user request initiates a complex decision process, where the platform must select, rank, and display a mixture of ads and organic content.
Traditionally, industrial advertising systems have relied on multi-stage cascading architecture, which is typically structured as: recall, pre-ranking, ranking, creative selection, and auction respectively. As shown in Figure~\ref{fig:pipline} (a), each stage selects the top-$k$ candidates from its input and forwards them to the subsequent stage. This paradigm effectively balances performance and efficiency by progressively narrowing down the optimal set of candidate items under strict latency constraints. 

However, it still suffers from intrinsic limitations. Earlier stages in the cascade inherently constrain the performance upper bound of downstream modules~\cite{deng2025onerec}. For example, ads filtered out in upstream recall or ranking stages cannot be recovered, even if they would be highly valuable in the final allocation, resulting in reduced platform revenue and failure to achieve global optimality. Despite various efforts to enhance overall recommendation performance by promoting interaction between ranking modules~\cite{gu2022ranking,zhang2023rethinking,zhao2023copr}, these approaches continue to operate within the traditional cascaded ranking framework.

The emergence of generative recommendation frameworks offers a new perspective on these longstanding challenges. Recent advances have demonstrated that transformer-based sequence models can unify retrieval, ranking, and generation in an end-to-end manner, delivering personalized content and capturing deep user-item dependencies~\cite{deng2025onerec,rajput2023recommender,yang2025sparse}. However, while such approaches have shown remarkable promise in organic recommendation scenarios, the complexity of industrial advertising presents a unique set of obstacles. Advertising systems must satisfy strict business constraints, including bidding, creative selection, ad slot allocation, and payment rules, while optimizing both user and platform objectives. These requirements introduce complex dependencies and practical challenges that cannot be fully addressed by directly applying existing generative recommendation models.

To address above challenges, we propose End-to-end Generative Advertising (EGA-V2), a novel generative framework that unifies all decision-making stages into a single model. EGA-V2 bridge the gap between generative modeling and the practical requirements of industrial advertising, which directly outputs the final ad sequence, as well as corresponding creatives, positions, and payment end-to-end. 
\textit{Firstly}, inspired by generative recommendation techniques, we leverage Residual Quantized Variational AutoEncoder (RQ-VAE) to encode user behavior and item features into hierarchical semantic tokens, and employ an encoder-decoder architecture with multi-token prediction to jointly generate candidate ad sequences and creative content.
\textit{Secondly}, we propose a token-level bidding and generative allocation mechanism, enhanced by permutation-aware reward modeling. This strategy decouples allocation from payment, allowing the model to effectively reflect business objectives during generation while approximately preserving incentive compatibility (IC) with a differentiable payment network.
\textit{Last but not least}, we introduce a multi-phase training paradigm. A pre-training phase learns user interests from actual exposure sequences that contain ad and organic items. Then auction-based post-training is applied to fine-tune the model with ad-specific constraints, dynamically optimizing ad allocation based on auction signals and platform objectives.

Our contributions are as follows:
\begin{itemize}[leftmargin=2em]
    \item We introduce EGA-V2, which surpasses traditional multi-stage cascading architectures by an unified single generative model. To the best of our knowledge, this is one of the first end-to-end generative advertising framework in industry.
    \item We propose a novel multi-phase training strategy consisting of interest-based pre-training and auction-based post-training. This design effectively balances user interests modeling with advertising-specific constraints, leverages diverse data sources during training, and aligns model outputs with final platform objectives.
    \item Extensive offline experiments in large-scale industrial datasets demonstrate the effectiveness and efficiency of our approach.
\end{itemize}

\section{Related Works}
\label{sec:related_works}
To achieve end-to-end optimization in industrial advertising, it is crucial to unify user behavior modeling, auction mechanisms, and ad allocation strategies. Motivated by this integration, we review recent advances in each area in this section.

\subsection{Generative Recommendation}
In recent years, there has been growing interest in applying generative paradigms to recommendation systems. A particularly promising direction is to formulate recommendation as a sequence generation task, where user-item interactions are modeled using transformer-based autoregressive architectures. These methods aim to deeply capture user behavioral context and generate personalized item sequences in an end-to-end fashion~\cite{de2020autoregressive,tang2023recent,rajput2023recommender,zhai2024actions,deng2025onerec,yang2025sparse,tay2022transformer}.

Tiger~\cite{rajput2023recommender} is a pioneering approach that introduces RQ-VAE to encode item content into hierarchical semantic IDs, allowing knowledge sharing across semantically similar items. Building upon this, COBRA~\cite{yang2025sparse} proposes a two-stage generation framework that first produces sparse IDs and then refines them into dense vectors, enabling a coarse-to-fine retrieval process. Another stream of research explores general-purpose generative recommenders. HSTU~\cite{zhai2024actions} reformulates recommendation as a sequential transduction task and designs a Transformer architecture tailored for high-cardinality, non-stationary streaming data. OneRec~\cite{deng2025onerec} further advances this by unifying retrieval, ranking, and generation within a single encoder-decoder framework, while incorporating session-level generation and preference alignment strategies to enhance output quality.
Besides, several methods focus on enhancing semantic tokenization and representation. LC-Rec~\cite{zheng2024adapting} aligns semantic IDs with collaborative filtering signals via auxiliary objectives. IDGenRec~\cite{tan2024idgenrec} leverages large language models to generate dense textual identifiers, showing strong generalization in zero-shot scenarios. SEATER~\cite{si2024generative} introduces tree-structured token spaces trained with contrastive and multi-task objectives to ensure consistency, while ColaRec~\cite{wang2024content} bridges content and interaction spaces for better alignment.

While these works demonstrate strong general recommendation performance, they are insufficient for online advertising systems, which require additional modeling of bidding, payment, and allocation constraints not captured in pure user interests modeling.

\subsection{Auction Mechanism}
Traditional auction mechanisms such as GSP~\cite{edelman2007internet} and its variants like uGSP~\cite{bachrach2014optimising} are widely deployed in online advertising due to their simplicity, interpretability, and strong revenue guarantees. However, they operate under the assumption of independence among ads, failing to account for externalities, that is, the influence of other ads~\cite{gatti2012truthful,hummel2014position}.

To address this, recent advances in computation have motivated the development of learning-based auction frameworks~\cite{zhang2021survey}. For example, DeepGSP~\cite{zhang2021optimizing} and DNA~\cite{liu2021neural} extend classical auctions by incorporating online feedback into end-to-end learning pipelines. However, DNA suffers from the evaluation-before-ranking dilemma, where the rank score must be predicted before knowing the final sequence, limiting its capacity to model set-level externalities.
Other approaches attempt to integrate optimality into auction design. NMA~\cite{liao2022nma} tackles this by exhaustively enumerating all possible allocations to ensure global optimality, but its computational cost renders it impractical for real-time applications. CGA~\cite{zhu2024contextual} addresses the limitations of traditional and learning-based ad auctions by explicitly modeling permutation-level externalities through an autoregressive allocation model and a gradient-friendly reformulation of incentive compatibility, enabling end-to-end optimization of both allocation and payment.

\subsection{Ad Allocation}
Platforms initially assigned fixed positions to ads and organic items, but dynamic ad allocation strategies are now gaining attention for their potential to optimize overall page-level performance~\cite{zhao2021dear,yan2020ads,xu2023multi,xie2021hierarchical,wang2022learning,liao2022deep}. CrossDQN~\cite{liao2022cross} introduces a DQN architecture to incorporate the arrangement signal into the allocation model without modifying the relative ranking of ads. HCA2E~\cite{chen2022hierarchically} proposes hierarchically constrained adaptive ad exposure that possesses the desirable game-theoretical properties and computational efficiency. MIAA~\cite{li2024deep} presents a deep automated mechanism that integrates ad auction and allocation, which simultaneously decides the ranking, payment, and display position of the ad. 

\section{Preliminary}
This section provides the necessary preliminaries for our approach. We define the core task in online advertising, and present the auction mechanism design that connects interests modeling with business objectives. The main notations are summarized in Table~\ref{tab:notation}.
\begin{table}[ht]
\centering
\caption{Summary of Notation}
\label{tab:notation}
\begin{tabular}{c|c}
\toprule
\textbf{Symbol} & \textbf{Description} \\
\midrule
$u$             & User $u$ \\
$X = \{x_1, \dots, x_N\}$ & Candidate set of $N$ ads \\
$O = \{o_1, \dots, o_M\}$ & Candidate set of $M$ organic contents \\
$\mathcal{Y} = (y_1, \dots, y_K)$   & Final ranked list of $K$ selected items \\
$\mathcal{S}^u=\{y_1,...,y_B\}$ & User historical behaviors of length $B$\\
$\mathcal{V}$   & Codebooks\\
$C$             & The number of codebook layers\\
$W$             & Codebook size of each layer\\
$\text{pCTR}_i$        & Predicted click-through rate of $i$-th item\\
$b_i$           & Bid value submitted by $i$-th advertiser \\
$v_i$           & Private value of $i$-th advertiser \\
$p_i$           & Payment charged to $i$-th advertiser\\
$u_i$           & Utility of $i$-th advertiser\\
$\bm{b}_{-i}$   & Bids profile of ads except $i$-th ad\\
$\mathcal{M} \langle \mathcal{A}, \mathcal{P} \rangle$ & Auction mechanism \\
$\text{Rev}$    & Platform expected revenue\\
$\bm{e}_i^{poi}$, $\bm{e}_i^{img}$      & POI and creative image embedding of item $i$\\
$\bm{a}_i^{poi}$, $\bm{a}_i^{img}$      & POI and creative image token of item $i$\\
$\mathcal{F}, \mathcal{R}$  & Pre-training model $\mathcal{F}$ and reward model $\mathcal{R}$\\
$P(\bm{a}_i)$   &  Probability of generating token $\bm{a}_i$\\
$\hat{r}$       & Estimated reward by reward model $\mathcal{R}$\\
\bottomrule
\end{tabular}
\end{table}

\subsection{Task Formulation}
We formalize a typical task of joint ad generation and allocation in online advertising systems. Given a page view (PV) request from user $u$, there are $N$ candidate ads and $M$ candidate organic items (non-sponsored). The organic sequence $O$ is assumed to be pre-ranked by an upstream module based on estimated GMV, and its internal order will remain fixed\footnote{This reflects common platform design constraints where organic content is ranked independently and must preserve user experience.}.
The system selects a final ranked list of $K$ items ($K \ll (N + M)$) to display:
\begin{equation}
    \mathcal{Y} = \{y_1, y_2, \dots, y_K\}, \quad y_i \in X \cup O,
\end{equation}
where the output list $\mathcal{Y}$ is generated by making the following decisions jointly under both user engagement and business constraints:
\begin{itemize}[leftmargin=2em]
    \item \textbf{Ad ranking:} decide the permutation of selected ads from $X$;
    \item \textbf{Creative selection:} for each chosen ad, generate the most appropriate creative image;
    \item \textbf{Payment:} compute the payment $p_i$ for each exposed ad based on its bid $b_i$ and allocated position;
    \item \textbf{Ad slot:} determine the optimal display position for each ad when completing with organic contents.
\end{itemize}
Each advertiser $i$ submits a bid $b_i$ corresponding to its private click value $v_i$. 
Our objective is maximize the expected platform revenue with sequence $\mathcal{Y}$:
\begin{equation}
    \max_{\theta} \; \mathbb{E}_{\mathcal{Y}}\left[\text{Rev}\right] = \max_{\theta} \; \mathbb{E}_{\mathcal{Y}} \left( \sum_{i=1}^{K} p_i \cdot \text{pCTR}_i \right).
    \label{eqn:obj}
\end{equation}

\subsection{Auction Mechanism Design}
\subsubsection{Auction Constraints}
Unlike traditional recommender systems, advertising platforms must not only optimize platform revenue but also ensure advertiser utility. Given an auction mechanism $\mathcal{M} \langle \mathcal{A}, \mathcal{P} \rangle$ with allocation rule $\mathcal{A}$ and payment rule $\mathcal{P}$, the expected utility $u_i$ for an advertiser $i$ is defined as:
\begin{equation}
    u_i(v_i; \bm{b}) = (v_i - p_i) \cdot \text{pCTR}_i,
\end{equation}

Two key economic constraints in auction mechanism must be satisfied in Equation~\eqref{eqn:obj}: incentive compatibility (IC) and individual rationality (IR).
IC requires that truthful bidding maximizes the advertiser's utility. For ad $x_i$, it holds that
\begin{equation}
    u_i(v_i; v_i, \bm{b}_{-i}) \geq u_i(v_i; b_i, \bm{b}_{-i}), \quad \forall v_i, b_i \in \mathbb{R}^+.
\end{equation}
IR requires that no advertiser pays more than their bid, that is, 
\begin{equation}
    p_i \leq b_i, \quad \forall i \in [N].
\end{equation}

\begin{figure*}[ht]
	\centering
	\includegraphics[width=0.9\linewidth,angle=0]{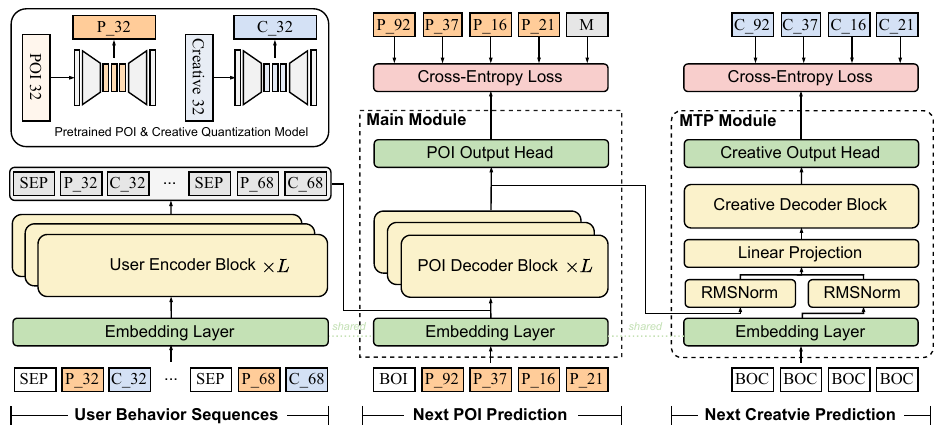}
	\caption{Overview of the interest-based pre-training architecture. The pre-training model consists of encoder for modeling historical behavior sequences, followed by a two-stage decoder: a POI decoder for next POI token prediction and a creative decoder for multi-modal next creative prediction. Both decoders are trained jointly using cross-entropy loss in MTP framework.}
	\label{fig:pretrain-model}
\end{figure*}

\subsubsection{Learning-based Auction}
To ensure incentive compatibility (IC) in our model, we adopt the concept of \textit{ex-post regret}~\cite{dutting2019optimal,zhu2024contextual} to quantify the potential gain an advertiser could obtain by untruthfully reporting their bid. This formulation enables us to enforce IC constraints in a differentiable manner, suitable for end-to-end optimization.

Formally, given the generated sequence $\mathcal{Y}$, ad $x_i\in\mathcal{Y}$ with true valuation $v_i$ and contextual user input $u$, the ex-post regret is defined as:
\begin{equation}
    \text{rgt}_i(v_i, \mathcal{Y}, u) = \max_{b^\prime_i} \left\{ u_i(v_i; b^\prime_i, \bm{b}_{-i}, \mathcal{Y}, u) - u_i(v_i; b_i, \bm{b}_{-i}, \mathcal{Y}, u) \right\},
\end{equation}
where $b_i$ is the truthful bid, $b'_i$ is a potential misreport, and $\bm{b}_{-i}$ represents bids excluding the item $x_i$.
The IC constraint is satisfied if and only if $\text{rgt}_i = 0$ for all advertisers. In practice, we approximate this using $N_v$ sampled valuations from distribution $\mathbb{F}$, the empirical ex-post regret for ad $x_i$ is
\begin{equation}
    \widehat{\operatorname{rgt}}_i = \frac{1}{N_v} \sum_{j=1}^{N_v} \text{rgt}_i(v_i^j, \mathcal{Y}, u).
    \label{eq:empirical_regret}
\end{equation}
We then formulate the auction design problem as minimizing the expected negative revenue under the constraint that the empirical ex-post regret remains zero for each ad $x_i$:
\begin{equation}
    \min_{\bm{w}} \; - \mathbb{E}_{\bm{v} \sim \mathbb{F}} \left[ \operatorname{Rev}^{\mathcal{M}} \right], \quad \text{s.t.} \quad \widehat{\text{rgt}}_i = 0, \; \forall i \in [N],
    \label{eq:ic_constrained_opt}
\end{equation}

\section{Methodology}
We introduce EGA-V2, an end-to-end generative advertising framework. This section describes how we learn user interests via vector tokenization and encoder-decoder design, leverage permutation-aware reward modeling for business objectives, incorporate auction-based preference alignment, and integrate all components through a unified multi-phase optimization strategy.
 

\subsection{Interest-based Pre-training}
\label{sec:pre-train}
The goal of the pre-training stage is to capture user interests based on their full historical behaviors, including both ads and organic contents. This stage serves as the foundation for learning a unified representation that reflects the interests of the users. Mathematically, the objective of the pre-training model $\mathcal{F}$ is to generate a interest-aware output sequence $\mathcal{Y}$ conditioned on the input user behavior sequence $\mathcal{S}^u$:
\begin{equation}
    \mathcal{Y} := \mathcal{F}(\mathcal{S}^u).
\end{equation}

\subsubsection{Feature Representations}
We represent the user-side context using $\mathcal{S}^u = \{y_1, y_2, \dots, y_B\}$, where each $y_i$ is a previously interacted item (e.g., click, purchase, like), and $B$ is the sequence length. The target label $\mathcal{Y} = \{y_1, y_2, \dots, y_K\}$ corresponds to high-value items actually exposed in the current PV session. Each candidate item $y_i$ is described by a multi-modal representation that includes Point of Interest (POI) feature embeddings $\bm{e}_i^{poi}$ (e.g., sparse ID features and dense features), and creative image embeddings $\bm{e}_i^{img}$ extracted from visual content. The final input can be represented as $\mathcal{S}^u = \{(\bm{e}_1^{poi},\bm{e}_1^{img}),(\bm{e}_2^{poi},\bm{e}_2^{img}),...,(\bm{e}_B^{poi},\bm{e}_B^{img})\}$.

\subsubsection{Vector Tokenization}
\label{sec:rq-vae}
Inspired by existing generative recommendation models~\cite{liu2024mmgrec,deng2025onerec,rajput2023recommender,yang2025sparse}, we employ Residual Quantized Variational Autoencoder (RQ-VAE)~\cite{zeghidour2021soundstream} to encode dense embeddings into semantic tokens. Each user behavior in the historical sequence is represented as a POI-creative pair:
\begin{equation}
    \mathcal{S}^u = \{(\bm{a}_1^{poi},\bm{a}_1^{img}), (\bm{a}_2^{poi},\bm{a}_2^{img}),...,(\bm{a}_B^{poi},\bm{a}_B^{img})\}.
\end{equation}
and the prediction target is a similarly structured sequence:
\begin{equation}
    \mathcal{Y} = \{(\bm{a}_1^{poi},\bm{a}_1^{img}), (\bm{a}_2^{poi},\bm{a}_2^{img}),...,(\bm{a}_K^{poi},\bm{a}_K^{img})\}.
\end{equation}
Each pair of tokens $(\bm{a}_i^{poi}, \bm{a}_i^{img})$ combines high-level categorical intent and fine-grained visual semantics. For simplicity, we assume that both POI and creative image are each represented by a single-level token $\bm{a}$, though the framework can be extended to hierarchical token representations if needed. For example, given the codebooks $\mathcal{V}$ that consists of $C$ layers, each of size $K$, the token $\bm{a}_i$ is denoted as 
\begin{equation}
    \bm{a}_i = (a_i^1, a_i^2, ..., a_i^C),\quad a_i^j\in \{\mathcal{V}_{j,1}, \mathcal{V}_{j,2},..., \mathcal{V}_{j,K}\}.
    \label{eq:token}
\end{equation}

\subsubsection{Probabilistic Decomposition}
The modeling of the target item's probability distribution is decomposed into two stages, leveraging the complementary advantages of POI-level and creative-level representations. Rather than directly predicting the next item $\bm{a}_{t+1}$ from the historical interaction sequence $\mathcal{S}^u_{1:t}$, EGA-V2 first predicts the POI $\bm{a}_{t+1}^{poi}$, then determines the creative image $\bm{a}_{t+1}^{img}$.
\begin{equation}
    P(\bm{a}_{t+1}^{poi}, \bm{a}_{t+1}^{img} \mid \mathcal{S}^u_{1:t}) = P(\bm{a}_{t+1}^{poi} \mid \mathcal{S}^u_{1:t})\cdot P(\bm{a}_{t+1}^{img} \mid \bm{a}_{t+1}^{poi}, \mathcal{S}^u_{1:t})
\end{equation}
where $P(\bm{a}_{t+1}^{poi} \mid S^u_{1:t})$ denotes the probability of generating the next POI $\bm{a}_{t+1}^{poi}$ based on the historical sequence $S^u_{1:t}$, capturing the categorical identity of the next item. Meanwhile, $P(\bm{a}_{t+1}^{img} \mid \bm{a}_{t+1}^{poi}, \mathcal{S}^u_{1:t})$ models the probability of generating the creative image $\bm{a}_{t+1}^{img}$ conditioned on the POI and the history, capturing fine-grained multi-modal details of the item.

\begin{figure*}[ht]
	\centering
	\includegraphics[width=0.9\linewidth,angle=0]{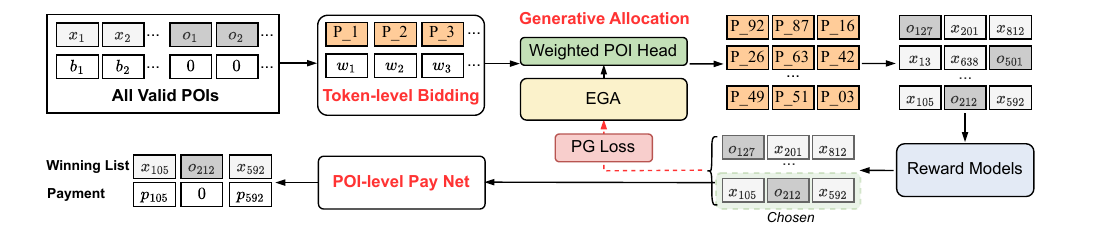}
	\caption{Illustration of the proposed generative ad allocation and payment architecture. Token-level bidding aggregates item bids for generative allocation, while a dedicated POI-level payment network ensures incentive-compatible (IC) constraint. The integrated framework supports dynamic trade-off between revenue and user experience.}
	\label{fig:auction-model}
\end{figure*}

\subsubsection{Encoder-decoder}
The overall generative architecture follows an encoder-decoder design aligned with the modular blocks shown in Figure~\ref{fig:pretrain-model}. The encoder module first encodes the user interaction sequence $\mathcal{S}^u$ using stacked self-attention and feed-forward layers:
\begin{equation}
    \mathcal{S}^e = \text{Encoder}(\mathcal{S}^u),
\end{equation}
where $\mathcal{S}^e$ represents the latent contextualized embedding of user interests.

The decoder generates the target sequence $\mathcal{Y}$ in an autoregressive manner, conditioned on the encoded user context $\mathcal{S}^e$. Each decoding step $t$ consists of two sub-stages: POI token generation followed by creative token generation. To enable joint learning, inspired by Multi-Token Prediction (MTP)~\cite{gloeckle2024better,liu2024deepseek}, we integrate a MTP module that supervises both heads simultaneously using cross-entropy loss on POI and creative predictions. All modules share the same token embedding space and quantization backbone, facilitating efficient training and consistent generation quality.

At step $t+1$, the POI decoder predicts the next POI token based on previously generated tokens and the encoded historical sequences:
\begin{equation}
    \bm{a}_{t+1}^{poi} = \text{POI-Decoder}(\mathcal{Y}_{1:t}, \mathcal{S}^e).
\end{equation}
Then, the creative decoder predicts the corresponding creative token by attending to both the POI prediction and the encoded context:
\begin{equation}
    \bm{a}_{t+1}^{img} = \text{Creative-Decoder}(\mathcal{Y}_{1:t}, \bm{a}_{t+1}^{poi}, \mathcal{S}^e).
\end{equation}

The generation of the entire target sequence $\mathcal{Y}$ is modeled as a product of conditional probabilities:
\begin{equation}
\label{eqn:joint_prob}
    P(\mathcal{Y} \mid \mathcal{S}^u) = \prod_{t=1}^{T} P(\bm{a}_{t}^{poi} \mid \mathcal{Y}_{1:t-1}, \mathcal{S}^e) \cdot P(\bm{a}_{t}^{img} \mid \mathcal{Y}_{1:t-1}, \bm{a}_t^{poi}, \mathcal{S}^e).
\end{equation}

\subsection{Permutation-aware Reward Model}
\label{sec:rm}
To ensure that the generated ad sequences align with real user interests, we introduce a permutation-aware reward model (RM) to guide iterative optimization. Unlike NLP tasks where interests signals are typically annotated by humans, the advertising domain benefits from more accurate, point-wise feedback derived from user interactions such as clicks and conversions.

Let $R(\mathcal{Y})$ denote the reward model that estimates reward signals for a candidate target items $\mathcal{Y} = \{\bm{a}_1, \bm{a}_2, \dots, \bm{a}_K\}$, where each $\bm{a}_i$ is a generated token. Besides, each generated token does not necessarily correspond to a unique item, which complicates the assignment of supervision signals in reward model.
To address this, we enrich each token representation $\bm{a}_i$ by concatenating raw item embeddings $\bm{e}_i^{poi}$, which is represented as:
\begin{equation}
    \bm{h}_i = [\bm{a}_i; \bm{e}_i^{poi}].
\end{equation}
where $[\cdot;\cdot]$ denotes concatenation. The target items becomes $\bm{h} = \{\bm{h}_1,\bm{h}_2, \dots, \bm{h}_K\}$.

The target items $\bm{h}$ are then processed by self-attention layers, which enable interaction among them to capture contextual dependencies and aggregate relevant information across the sequence.
\begin{equation}
    \bm{h}_f = \text{SelfAttention}(\bm{h}W^Q, \bm{h}W^K, \bm{h}W^V).
\end{equation}

For making fine-grained predictions such as pCTR for POI and creative image respectively, and pCVR, we augment the reward model with multiple task-specific towers:
\begin{equation}
\label{eqn:listwise_pcxr}
    \hat{r}^{\text{pctr}} = \text{Tower}^{\text{pctr}} \left( \sum \bm{h}_f \right), \quad 
    \hat{r}^{\text{pcvr}} = \text{Tower}^{\text{pcvr}} \left( \sum \bm{h}_f \right)
\end{equation}
where $\text{Tower}(\cdot) = \text{Sigmoid}(\text{MLP}(\cdot))$. After obtaining the estimated rewards $\hat{r}^{\text{pcxr}}$ and the corresponding ground-truth labels $y^{\text{pcxr}}$ for each item and reward type, we train the reward model by directly minimizing the binary cross-entropy loss. The detailed training procedure is described in the Section~\ref{sec:training}.

\subsection{Auction-based Preference Alignment}
\label{sec:align}
In the industrial advertising scenario, generative recommendation models need to fulfill not only user interests but also critical business constraints. Specifically, two main business constraints must be simultaneously satisfied: i) advertiser demands for effective exposure, bidding compatibility, and payment consistency; and ii) platform constraints for balancing revenue with user experience, such as controlling ad exposure ratios. To address these challenges, we propose an integrated generative allocation and payment strategy to do preference alignment, consisting of a token-level bidding based allocation module and a decoupled POI-level payment network. The overall structure is illustrated in Figure~\ref{fig:auction-model}.

\subsubsection{Token-level Bidding}
After applying RQ-VAE for item tokenization, the original item-level bids no longer align directly with the token space due to a many-to-many mapping between items and latent tokens. Inspired by Google's token-level bidding theory~\cite{duetting2024mechanism}, we introduce a token-level allocation mechanism that aggregates bids across items associated with each token. As mentioned in Equation~\eqref{eq:token}, each token $\bm{a}_i=(a_i^1,a_i^2,...,a_i^C)$. To maintain differentiation in bids across different tokens, the bid of token $a_i^j$ is $b(a_i^j) = \max (b_1,b_2,...,b_{N_i})$, where $N_i$ is the number of items associated with token $a_i^j$.
Formally, the aggregated bid weight for token $a_i^j$ is defined as:
\begin{align}
w(a_i^j) = \left[b(a_i^j)\right]^\alpha + \beta = \left[\max (b_1,b_2,...,b_{N_i})\right]^\alpha + \beta
\label{eq:token_bid}
\end{align}
where $\alpha$ and $\beta$ are hyperparameters. By adjusting $\alpha$, the influence of bids on the allocation process can be flexibly controlled in real-time; meanwhile, dynamically tuning $\beta$ enables effective balancing of the proportion between ads and organic content in the generated results.

\subsubsection{Generative Allocation}
The probability of selecting token $a_i^j$ in the generative allocation is given by a softmax normalization:
\begin{equation}
\label{eqn:allocation_prob}
z(a_i^j) = \frac{w(a_i^j)\cdot e^{a_i^j}}{\sum_{k=1}^W\left[w(a^{j,k})\cdot e^{a^{j,k}}\right]},
\end{equation}
where $a^{j,k}$ is the $k$-th code of $j$-th layer.
Based on the generative probabilities $\bm{z}\in \mathbb{R}^{C\times W}$, where $C$ is the codebook layers and $W$ is the codebook size of each layer, we apply beam search to generate $N_S$ candidate sequences of length $K$, ensuring both diversity and high-quality selection in the allocation:
\begin{equation}
{\mathcal{S}^{(1)}, \mathcal{S}^{(2)}, \dots, \mathcal{S}^{(N_S)}} = \text{BeamSearch}(\bm{z}, N_S),
\end{equation}
where $\mathcal{S}^{(j)}$ represents $j$-th generated candidate sequence of tokens, and $N_S$ denotes both the beam width and the number of generated candidate sequences.
We evaluate each sequence using the reward model $\mathcal{R}$ to measure its expected business value, which outputs the reward $\hat{r}_j$ for $j$-th sequence:
\begin{equation}
    \hat{r}_j = \mathcal{R}(S^{(j)}).
\end{equation}
The final output is the sequence $\mathcal{S}^*$ with the highest reward. It is worth noting that RM provides a flexible interface to accommodate diverse reward signal combinations as needed by the platform.

\subsubsection{POI-level Payment Network}
While allocation operates at the token level, payment is calculated at the item or POI level, which aligns better with traditional advertiser expectations and business logics. The decoupled payment network is specifically designed to satisfy IC and IR constraints, which computes payments based on item representations. 

Formally, the payment network inputs include item representations $\mathcal{S}^*=\{y_1,y_2,...,y_K\}\in \mathbb{R}^{K\times d}$, the self-exclusion bidding profile $\mathcal{B^-}=\{\bm{b}_{-1},\bm{b}_{-2},...,\bm{b}_{-K}\}\in \mathbb{R}^{K\times (K-1)}$, and the expected value profile $\mathcal{Z}\cdot \Theta \in \mathbb{R}^{K\times 1}$, where $\mathcal{Z}=\{z_1,z_2,...,z_K\}$ denotes the allocation probability defined in Equation~\eqref{eqn:allocation_prob} and $\Theta=\{\hat{r}_1^{pctr},\hat{r}_2^{pctr},...,\hat{r}_K^{pctr}\}$ denotes the permutation-aware pCTR estimated by the reward model in Equation~\eqref{eqn:listwise_pcxr}. The payment rate is defined as:
\begin{equation}
\hat{p} = \sigma(\text{MLP}(\mathcal{S}^*;\mathcal{B^-};\mathcal{Z} \cdot \Theta)) \in [0,1]^K,
\end{equation}
where $\sigma$ denotes the sigmoid activation to satisfy IR constraint, and the final POI-level payment $p$ is calculated as:
\begin{equation}
p = \hat{p}\odot b.
\end{equation}
It should be noted that payments are calculated exclusively for ads if $y_i\in X$, while the payment for organic content is zero if $y_i \in O$.

\subsection{Optimization and Training}
\label{sec:training}
\begin{figure}[ht]
	\centering
	\includegraphics[width=\linewidth,angle=0]{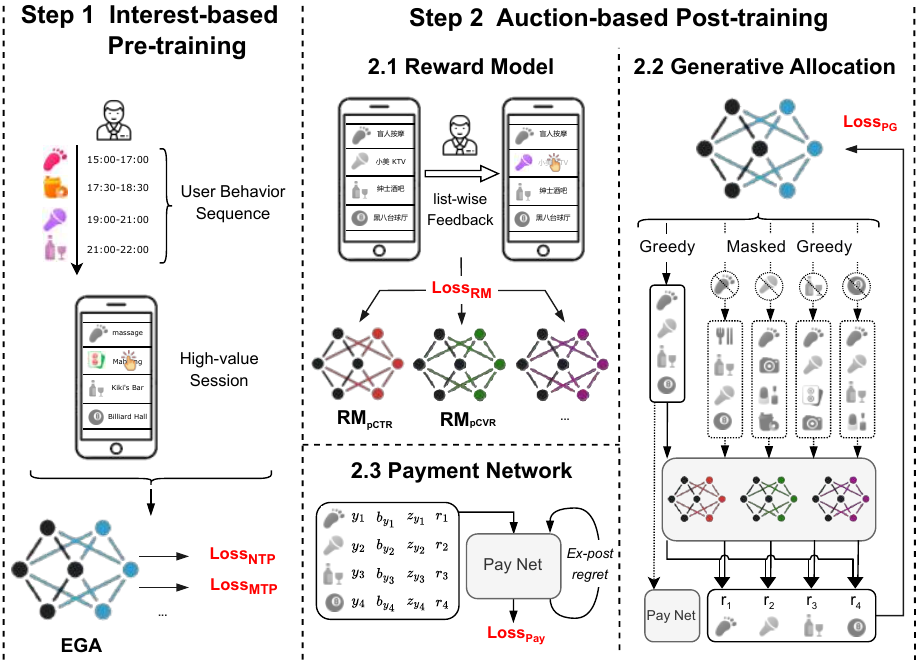}
	\caption{Overview of the proposed optimization and training pipeline. The training process consists of two steps: 1) Interest-based Pre-training, where behavior sequences are used to pre-train EGA-V2 that predicts the next POI and creative tokens; and 2) Auction-based Post-training, which includes: 2.1) a permutation-aware reward model trained with list-wise feedback to predict reward signals, such as pCTR and pCVR; 2.2) generative allocation, where different candidate sequences are scored by the reward model to optimize policy gradients; and 2.3) a payment network optimized with ex-post regret to ensure IC approximately.}
	\label{fig:training}
\end{figure}
The overall procedure of training is shown in Figure~\ref{fig:training}.
\subsubsection{Interest-based Pre-training}
In the pre-training phase, we first train the generative backbone to capture user interests from historical interaction behaviors. Specifically, we optimize two separate cross-entropy losses for predicted target sequence $\mathcal{Y}$: next-POI prediction loss $\mathcal{L}_{\text{NTP}}$ of main module, and next-creative prediction loss $\mathcal{L}_{\text{MTP}}$ of MTP module. Formally, according to Equation~\eqref{eqn:joint_prob},
\begin{align}
    \mathcal{L}_{\text{NTP}} &= -\frac{1}{K}\sum_{i=1}^K \log P(\bm{a}_i^{poi} \mid \mathcal{Y}_{1:i-1},\mathcal{S}^e)\\
    \mathcal{L}_{\text{MTP}} &= -\frac{1}{K}\sum_{i=2}^{K+1}\log P(\bm{a}_{i-1}^{img} \mid \mathcal{Y}_{1:t-1}, \bm{a}_{i}^{poi},\mathcal{S}^e)
\end{align}
Then total pre-training loss is defined as:
\begin{equation}
\mathcal{L}_{\text{pre-train}} = \mathcal{L}_{\text{NTP}} + \mathcal{L}_{\text{MTP}}.
\end{equation}

\subsubsection{Auction-based Post-training}
Following the pre-training phase, we freeze its parameters and optimize the generative advertising model under auction constraints through an auction-based post-training stage. This phase aligns the generative outputs with platform revenue objectives and advertiser demands, consisting of two components: i) reward model training, ii) generative allocation training based on policy gradient, and iii) payment network optimization based on Lagrangian method.

\textbf{Reward Model Training.} We train a separate reward model (RM) using users' real feedback signals (e.g., clicks and conversions). The RM is optimized by minimizing the binary cross-entropy loss:
\begin{equation}
\mathcal{L}_{\text{RM}}^{\text{pcxr}}\! =\! -\frac{1}{|\mathcal{D}|}\! \sum_{d\in\mathcal{D}}\! \sum_{i=1}^K \left( y^{\text{pcxr}} \log \hat{r}^{\text{pcxr}} + (1 - y^{\text{pcxr}}) \log (1 - \hat{r}^{\text{pcxr}}) \right),
\end{equation}
where $y^{\text{pcxr}}$ represents ground-truth labels derived from real user interactions, $\hat{r}^{\text{pcxr}}$ is the predicted probability from the reward model by Equation~\eqref{eqn:listwise_pcxr}, and $\mathcal{D}$ is the training dataset.

\textbf{Generative Allocation Training.} After convergence of the reward model, we adopt a non-autoregressive policy gradient based method. Given a generated winning ad sequence $\mathcal{S}^* = \{y_1,y_2,...,y_K\}$, we define the marginal contribution of each item $y_i$ to platform revenue as:
\begin{equation}
r_{y_i} = \sum_{y_j \in \mathcal{S}^*} b_j \hat{r}^{\text{pctr}}_j - \sum_{y_j \in \mathcal{S}^*_{-i}} b_j \hat{r}^{\text{pctr}}_j,
\end{equation}
where $\mathcal{S}^*_{-i}$ denotes the best alternative ad sequence excluding $y_i$.
We then apply a policy gradient objective to maximize the expected rewards:
\begin{equation}
\mathcal{L}_{\text{PG}} = -\frac{1}{|\mathcal{D}|} \sum_{d\in\mathcal{D}} \sum_{y_i\in \mathcal{S}^*} r_{y_i} \log z_{y_i},
\end{equation}
where $z_{y_i}$ is the allocation probability for item $y_i$ by Equation~\eqref{eqn:allocation_prob}. This design encourages the generator to produce sequences that yield higher overall revenue, using fixed reward model parameters.

\textbf{Payment Network Optimization}. The payment network optimizes Equation~\eqref{eq:ic_constrained_opt} to balance revenue maximization and IC constraint via Lagrangian dual formulation. The loss function integrates both total platform payment and ex-post regret minimization. Given the selected sequence $\mathcal{Y}$, the payment loss $\mathcal{L}_{\text{Pay}}$ is defined as follows:
\begin{equation}
    \mathcal{L}_{\text{Pay}} \! =\! -\frac{1}{|\mathcal{D}|}\sum_{d \in \mathcal{D}}\! \left(
    \sum_{y_i \in \mathcal{S}^*} p_i \hat{r}^{\text{pctr}}_i \!-\! \sum_{y_i \in \mathcal{S}^*}\! \lambda_i \widehat{\text{rgt}}_i\! -\! \frac{\rho}{2} \! \sum_{y_i \in \mathcal{S}^*}\! (\widehat{\text{rgt}}_i)^2
    \right),
\end{equation}
where $\widehat{\text{rgt}}_i$ is the ex-post regret of ad $y_i$, $p_i$ is the predicted payment from the payment network, $\lambda_{i}$ is a Lagrange multiplier, and $\rho>0$ is the hyperparameter for the IC penalty term.
To solve this constrained optimization, we adopt an iterative Lagrangian-based approach to jointly optimize the payment network. Specifically, we alternate between two steps:
\begin{itemize}[leftmargin=2em]
    \item Payment Network Update: Optimize the parameters $\bm{\theta}_{\text{Pay}}$ of the payment network by minimizing the Lagrangian objective with fixed multipliers:
    \begin{equation}
        \bm{\theta}_{\text{Pay}}^{\text{new}} = \arg\min_{\bm{\theta}_{\text{Pay}}} \; \mathcal{L}_{\text{Pay}}(\bm{\theta}_{\text{Pay}}^{\text{old}}, \bm{\lambda}^{\text{old}}).
    \end{equation}

    \item Multiplier Update: Adjust the Lagrange multipliers based on the observed empirical ex-post regret:
    \begin{equation}
        \bm{\lambda}^{\text{new}} = \bm{\lambda}^{\text{old}} + \rho \cdot \widehat{\operatorname{rgt}}(\bm{\theta}_{\text{Pay}}^{\text{new}}).
    \end{equation}
\end{itemize}

Note that the overall objective is non-convex, and convergence to the global optimum is not theoretically guaranteed. However, our empirical results show that this optimization strategy effectively minimizes regret while maintaining near-optimal revenue in real-world scenarios.

\section{Experiments}
In this section, we evaluate our proposed model on industrial dataset and aim to answer the following research questions\footnote{Due to computational complexity constraints, our current evaluation is limited to offline experiments, with online A/B testing reserved for future work.}:
\begin{itemize}[leftmargin=2em]
    \item \textbf{RQ1}: How does our EGA-V2 model perform, compared to the state-of-the-art advertising models?
    \item \textbf{RQ2}: What is the impact of designs (e.g. MTP module, token-level bidding, payment network, and multi-phase training) on the performance of EGA-V2?
    \item \textbf{RQ3}: How do hyperparameters affect model performance?
\end{itemize}

\subsection{Experiment Setup}
\subsubsection{Dataset}
The industrial dataset used in our experiments consists of real interaction logs collected from a large-scale location-based services (LBS) platform Meituan, spanning the period from September 2024 to April 2025. The dataset contains 200 million requests from over 2 million users and nearly 10 million unique ads of low-traffic ad slot in Meituan. We use the first 200 days for pre-training and randomly sample 10\% data for preference alignment. The last 14 days are used for testing.

\subsubsection{Evaluation Metrics}
In offline experiments, we employ the following metrics\footnote{To protect business confidentiality, the reported results on Meituan have been transformed in a way that preserves their statistical properties while ensuring that sensitive business information cannot be inferred or reconstructed from the published data.} to comprehensively evaluate platform revenue, user experience, and ex-post regret of advertisers, respectively.
\begin{itemize}[leftmargin=2em]
    \item \textbf{Revenue Per Mille:} $\text{RPM} = \frac{\sum \text{click} \times \text{payment}}{\sum \text{impression}} \times 1000$.
    \item \textbf{Click-Through Rate:} $\text{CTR} = \frac{\sum \text{click}}{\sum \text{impression}}$.
    \item \textbf{IC Metric:} $\Psi = \frac{1}{|\mathcal{D}|} \sum_{d \in \mathcal{D}} \sum_{i \in k} \frac{\widehat{\text{rgt}}_i^d}{u_i(v_i^d; \bm{b}^d)}$,
    where $\widehat{\text{rgt}}_i^d$ denotes the empirical ex-post regret for advertiser $i$ in session data $d$ as defined in Equation~\eqref{eq:empirical_regret}, and $u_i$ is the realized utility. This metric evaluates incentive compatibility (IC), representing the relative utility gain an advertiser could obtain by manipulating its bid~\cite{deng2020data,liao2022nma,liu2021neural}. Following~\cite{wang2022designing,zhu2024contextual}, IC is empirically tested via counterfactual perturbation: for each advertiser, the bid $b_i$ is replaced with $\gamma \times b_i$, where $\gamma \in \{0.2 \times j \mid j=1,2,\dots,10\}$.
\end{itemize}
For offline experiments, evaluation metrics are computed using the predicted values from the reward model.

\subsubsection{Baselines}
We evaluate EGA-V2 against the following two widely adopted industrial architectures.
\begin{itemize}[leftmargin=2em]
    \item \textbf{MCA:} The multi-stage cascading architecture (MCA) is a standard paradigm in industrial online advertising. It consists of five key stages: recall, ranking, creative selection, auction, and ad allocation. For a strong baseline, we implement MCA using representative methods: Tiger~\cite{rajput2023recommender} for recall, HSTU~\cite{zhai2024actions} for ranking, Peri-CR~\cite{yang2024parallel} for creative selection, CGA~\cite{zhu2024contextual} for auction, and CrossDQN~\cite{liao2022cross} for ad allocation.
    \item \textbf{GR:} Generative recommendation (GR) formulates recommendation as a sequence generation task, where user-item interactions are modeled using transformer based autoregressive architectures. To apply GR in online advertising scenario, we construct this baseline by integrating OneRec~\cite{deng2025onerec} with Peri-CR~\cite{yang2024parallel} for creative selection and GSP~\cite{edelman2007internet} for payment.
\end{itemize}

\subsubsection{Implementation Details}
We train EGA-V2 using the Adam optimizer with an initial learning rate of 0.0024. The batch size is set to 128. Model training and optimization are performed on NVIDIA A100 GPUs with 80G memory. For hyperparameters, we tried different hyperparameters using grid search. Due to space constraints, we report only the most optimal hyperparameter settings in this paper. For interest-based pre-training, the block number $L$ of encoder and decoder is 3, the number of codebook layers $C=3$, and codebook size $W$ of each layer is 1024. We consider $K=10$ target session items and use $B=256$ historical behaviors as context. For auction-based preference alignment, the hyperparameters in token-level bidding $\alpha=1.2$ and $\beta=2$. We generate $N_S=64$ different sequences for each request by beam search. For reward model and payment network, the hidden layers of the MLP are 128, 32, and 10.

\subsection{Offline Performance (RQ1)}
\begin{table}[ht]
\centering
\caption{The experimental results of out model EGA-V2 and competitors on industrial dataset. The bold value marks the best one in each column. Each result is presented in the form of mean (lift percentage). Lift percentage means the improvement of EGA-V2 over the best baselines.}
\label{tab:exp1}
\begin{tabular}{l|cccc}
\toprule
\textbf{Model} & \textbf{RPM} & \textbf{CTR-poi} & \textbf{CTR-img} & $\mathbf{\Psi}$ \\
\midrule
MCA & 192.45 (-16.5\%) & 0.0558 (-8.8\%) & 0.0529 (-9.3\%) & 3.6\%\\
GR  & 206.73 (-10.3\%) & 0.0582 (-4.9\%) & 0.0546 (-6.3\%) & 8.4\% \\
EGA-V2 & \textbf{230.41}  & \textbf{0.0612} & \textbf{0.0583} & \textbf{2.7\%}\\
\bottomrule
\end{tabular}
\end{table}

As shown in Table~\ref{tab:exp1}, our key observations are 1) the Multi-stage Cascading Architecture (MCA) suffers from the well-known early-stage filtering problem: promising ads are often eliminated in the initial recall or ranking stages, leading to suboptimal overall performance. This limitation is reflected in its relatively lower RPM and CTR metrics compared to more unified approaches. 2) In contrast, the generative recommendation baseline (GR) is designed to improve sequence modeling and personalization. However, GR still falls short in real advertising scenarios due to its limited ability to satisfy practical business constraints. Specifically, GR does not guarantee incentive compatibility with GSP (as evidenced by a much higher IC regret of 8.4\%), cannot flexibly implement dynamic ad allocation or control ad exposure rates, and is unable to support parallel creative selection for each ad. These limitations reduce the effectiveness of GR when applied to industrial advertising.

Our proposed EGA-V2 overcomes these shortcomings by integrating end-to-end generation, permutation-aware reward modeling, token-level bidding, and a dedicated payment network. As a result, EGA-V2 achieves the best overall performance: it significantly improves revenue (RPM), enhances both POI and creative CTRs, and delivers superior economic robustness with the lowest IC regret among all baselines. This demonstrates that addressing both early-stage candidate loss and business constraints is crucial for practical deployment in industrial advertising systems.

\subsection{Ablation Study (RQ2)}
\begin{table}[ht]
\centering
\caption{Ablation Study of EGA-V2.}
\label{tab:exp2}
\begin{tabular}{l|cccc}
\toprule
\textbf{Model} & \textbf{RPM} & \textbf{CTR-poi} & \textbf{CTR-img} & $\mathbf{\Psi}$ \\
\midrule
EGA-V2 & \textbf{230.41}     & \textbf{0.0612} & \textbf{0.0583} & \textbf{2.7\%}\\
EGA-mtp      & 225.45 (-2.1\%)   & 0.0593 (-3.1\%)   & 0.0562 (-3.6\%)   & 2.7\%   \\
EGA-end      & 218.21 (-5.3\%)   & 0.0600 (-2.0\%)   & 0.0572 (-1.9\%)  & 3.0\%   \\
EGA-bid      & 222.94 (-3.2\%)   & 0.0603 (-1.5\%)   & 0.0576 (-1.2\%)  & 4.1\%   \\
EGA-gsp      & 226.17 (-1.8\%)   & 0.0608 (-0.6\%)   & 0.0580 (-0.5\%)  & 8.2\%   \\
\bottomrule
\end{tabular}
\end{table}
We conduct experiments to validate the effectiveness of different components. Correspondingly, we design a series of ablation studies, which considers four variants to simplify EGA-V2 in different ways:
\begin{itemize}[leftmargin=2em]
    \item \textbf{EGA-mtp} removes the Multi-Token Prediction (MTP) module, replacing it with a standard next-token prediction (NTP) approach that independently predicts POI and creative in a sequential manner. This setting examines the importance of joint modeling for POI and creative generation.
    \item \textbf{EGA-end} disables the multi-phase training strategy and instead trains the entire model in a single end-to-end stage. The comparison assesses the benefit of our proposed interest-based pre-training and auction-based post-training pipeline. 
    \item \textbf{EGA-bid} simplifies the token-level bidding mechanism by replacing the maximum aggregation with an average operation, i.e., each token's bid is computed as the average over all relevant items rather than the maximum. Formally, $b(a_i^j)=\text{avg}(b_1,b_2,...,b_{N_i})$. This tests the effect of aggregation strategy in the bidding process.
    \item \textbf{EGA-gsp} removes the dedicated payment network and replaces it with standard GSP payment computation, keeping other modules unchanged. The comparison highlights the role of the learned payment network in optimizing revenue and enforcing incentive compatibility.
\end{itemize}

The performance results are shown in Table~\ref{tab:exp2}, from which we observe the full EGA-V2 model consistently outperforms all ablation variants across all major metrics, validating the effectiveness of our design. Specifically,
\begin{enumerate}[leftmargin=2em]
    \item EGA-mtp leads to a noticeable drop in RPM, CTR-poi, and CTR-img (-2.1\%, -3.1\%, and -3.6\% respectively), indicating the importance of jointly modeling POI and creative selection.
    \item The EGA-end variant, which disables multi-phase training, also results in a performance decrease, especially on RPM (-5.3\%), highlighting the benefit of our two-stage optimization strategy for aligning user interests modeling and business objectives.
    \item The EGA-bid variant, which replaces the max aggregation in token-level bidding with an average operation, causes a moderate decrease in all metrics, and notably increases the IC regret $\Psi$ to 4.1\%, showing the importance of our aggregation choice for incentive compatibility and allocation efficiency.
    \item EGA-gsp results in the highest IC regret ($\Psi$ rises from 2.7\% to 8.2\%) due to the absence of payment network and degrading to GSP payment, although other metrics drop only slightly. This demonstrates that the payment network is crucial for achieving incentive compatibility approximately in practice.
\end{enumerate}
In summary, each component of EGA-V2, including multi-token prediction, multi-phase training, max aggregation in bidding, and a dedicated payment network, plays an important and complementary role in improving revenue and user experience, and ensuring economic robustness.

\subsection{Hyperparameters (RQ3)}
\begin{figure}[!htbp]
	\begin{subfigure}[b]{0.5\linewidth}
		\centering
		\includegraphics[width=\linewidth]{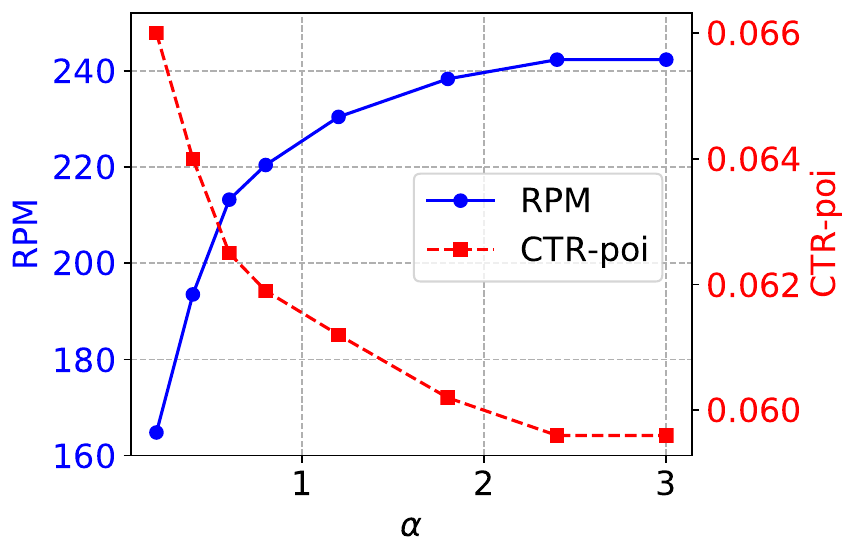}
		\label{fig:exp_alpha}
	\end{subfigure}%
	\begin{subfigure}[b]{0.5\linewidth}
		\centering
		\includegraphics[width=\linewidth]{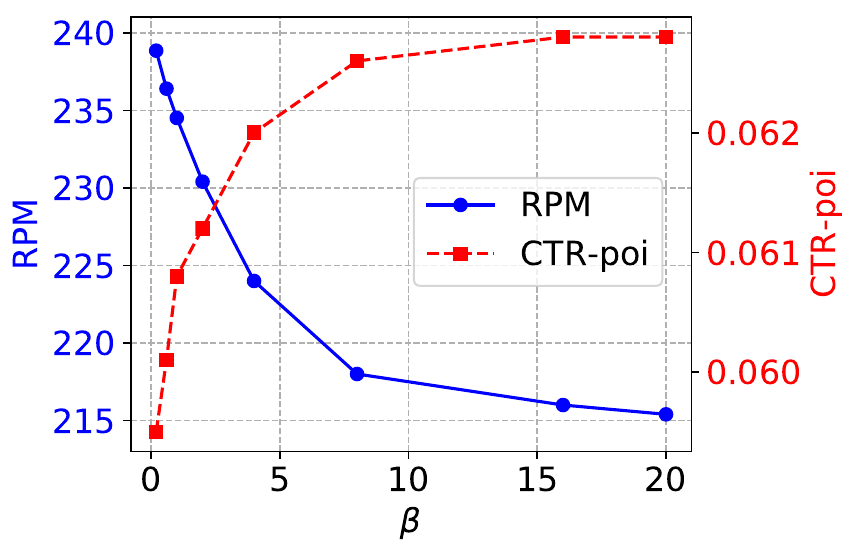}
		\label{fig:exp_beta}
	\end{subfigure}
        \centering
        \vspace{-10mm}
	\caption{Effect of hyperparameters of EGA-V2.}
	\label{fig:exp_params}
\end{figure}

We conduct a comprehensive study to evaluate the sensitivity of EGA-V2 with respect to key hyperparameters. The results are summarized in Figure~\ref{fig:exp_params}, where RPM and CTR-poi are used as the evaluation metrics.
Figure~\ref{fig:exp_params} examines the impact of token-level bidding parameters $\alpha$ and $\beta$ in Equation~\eqref{eq:token_bid}. $\alpha$ measures the importance of bids during the allocation process. Increasing $\alpha$ significantly boosts RPM at first, as higher-bid ads become more likely to win in the generative allocation. However, this trend gradually plateaus as further increasing $\alpha$ provides diminishing marginal returns. In contrast, CTR-poi shows a consistent decline as $\alpha$ rises, reflecting a shift in exposure from more organic contents to higher-bid ads. Besides, $\beta$ acts as a global balancing factor between ads and organic content. Increasing $\beta$ reduces the probability of ads being allocated. As a result, RPM decreases steadily with increasing $\beta$, and eventually flattens out as ads rarely win positions in the list. Conversely, CTR-poi increases with $\beta$, as the display list becomes dominated by organic content. In practice, the optimal values of $\alpha$ and $\beta$ are selected to achieve a trade-off between platform revenue and user experience.

\section{Conclusion}
\label{sec:conc}
In this work, we presented End-to-end Generative Advertising (EGA-V2), a novel framework that unifies ranking, creative selection, ad allocation, and payment computation into a single generative model for industrial advertising systems. By leveraging hierarchical semantic tokenization, permutation-aware reward modeling, and token-level bidding and allocation, EGA-V2 bridges the gap between user interests modeling and business-critical auction constraints such as IC and IR. The proposed multi-phase training paradigm, including interest-based pre-training and auction-based post-training, ensures that EGA-V2 captures both user interests and advertiser utility under complex real-world conditions.
Extensive offline experiments on an industrial dataset demonstrate that EGA-V2 achieves substantial improvements over both multi-stage cascading architecture and recent generative recommendation baselines in platform revenue, user experience, and advertiser return on investment. Our ablation studies validate the effectiveness of each design component and highlight the flexibility of the proposed framework. 

Looking forward, we believe EGA-V2 opens new directions for the integration of generative modeling and economic mechanism design in online advertising. Future work will further explore scaling laws and enhance business interpretability.



\bibliographystyle{ACM-Reference-Format}
\bibliography{ref}


\begin{thebibliography}{41}


\ifx \showCODEN    \undefined \def \showCODEN     #1{\unskip}     \fi
\ifx \showISBNx    \undefined \def \showISBNx     #1{\unskip}     \fi
\ifx \showISBNxiii \undefined \def \showISBNxiii  #1{\unskip}     \fi
\ifx \showISSN     \undefined \def \showISSN      #1{\unskip}     \fi
\ifx \showLCCN     \undefined \def \showLCCN      #1{\unskip}     \fi
\ifx \shownote     \undefined \def \shownote      #1{#1}          \fi
\ifx \showarticletitle \undefined \def \showarticletitle #1{#1}   \fi
\ifx \showURL      \undefined \def \showURL       {\relax}        \fi
\providecommand\bibfield[2]{#2}
\providecommand\bibinfo[2]{#2}
\providecommand\natexlab[1]{#1}
\providecommand\showeprint[2][]{arXiv:#2}

\bibitem[Bachrach et~al\mbox{.}(2014)]%
        {bachrach2014optimising}
\bibfield{author}{\bibinfo{person}{Yoram Bachrach}, \bibinfo{person}{Sofia Ceppi}, \bibinfo{person}{Ian~A Kash}, \bibinfo{person}{Peter Key}, {and} \bibinfo{person}{David Kurokawa}.} \bibinfo{year}{2014}\natexlab{}.
\newblock \showarticletitle{Optimising trade-offs among stakeholders in ad auctions}. In \bibinfo{booktitle}{\emph{Proceedings of the fifteenth ACM conference on Economics and computation}}. \bibinfo{pages}{75--92}.
\newblock


\bibitem[Chen et~al\mbox{.}(2022)]%
        {chen2022hierarchically}
\bibfield{author}{\bibinfo{person}{Dagui Chen}, \bibinfo{person}{Qi Yan}, \bibinfo{person}{Chunjie Chen}, \bibinfo{person}{Zhenzhe Zheng}, \bibinfo{person}{Yangsu Liu}, \bibinfo{person}{Zhenjia Ma}, \bibinfo{person}{Chuan Yu}, \bibinfo{person}{Jian Xu}, {and} \bibinfo{person}{Bo Zheng}.} \bibinfo{year}{2022}\natexlab{}.
\newblock \showarticletitle{Hierarchically constrained adaptive ad exposure in feeds}. In \bibinfo{booktitle}{\emph{Proceedings of the 31st ACM International Conference on Information \& Knowledge Management}}. \bibinfo{pages}{3003--3012}.
\newblock


\bibitem[De~Cao et~al\mbox{.}(2020)]%
        {de2020autoregressive}
\bibfield{author}{\bibinfo{person}{Nicola De~Cao}, \bibinfo{person}{Gautier Izacard}, \bibinfo{person}{Sebastian Riedel}, {and} \bibinfo{person}{Fabio Petroni}.} \bibinfo{year}{2020}\natexlab{}.
\newblock \showarticletitle{Autoregressive entity retrieval}.
\newblock \bibinfo{journal}{\emph{arXiv preprint arXiv:2010.00904}} (\bibinfo{year}{2020}).
\newblock


\bibitem[Deng et~al\mbox{.}(2025)]%
        {deng2025onerec}
\bibfield{author}{\bibinfo{person}{Jiaxin Deng}, \bibinfo{person}{Shiyao Wang}, \bibinfo{person}{Kuo Cai}, \bibinfo{person}{Lejian Ren}, \bibinfo{person}{Qigen Hu}, \bibinfo{person}{Weifeng Ding}, \bibinfo{person}{Qiang Luo}, {and} \bibinfo{person}{Guorui Zhou}.} \bibinfo{year}{2025}\natexlab{}.
\newblock \showarticletitle{OneRec: Unifying Retrieve and Rank with Generative Recommender and Iterative Preference Alignment}.
\newblock \bibinfo{journal}{\emph{arXiv preprint arXiv:2502.18965}} (\bibinfo{year}{2025}).
\newblock


\bibitem[Deng et~al\mbox{.}(2020)]%
        {deng2020data}
\bibfield{author}{\bibinfo{person}{Yuan Deng}, \bibinfo{person}{S{\'e}bastien Lahaie}, \bibinfo{person}{Vahab Mirrokni}, {and} \bibinfo{person}{Song Zuo}.} \bibinfo{year}{2020}\natexlab{}.
\newblock \showarticletitle{A data-driven metric of incentive compatibility}. In \bibinfo{booktitle}{\emph{Proceedings of The Web Conference 2020}}. \bibinfo{pages}{1796--1806}.
\newblock


\bibitem[Duetting et~al\mbox{.}(2024)]%
        {duetting2024mechanism}
\bibfield{author}{\bibinfo{person}{Paul Duetting}, \bibinfo{person}{Vahab Mirrokni}, \bibinfo{person}{Renato Paes~Leme}, \bibinfo{person}{Haifeng Xu}, {and} \bibinfo{person}{Song Zuo}.} \bibinfo{year}{2024}\natexlab{}.
\newblock \showarticletitle{Mechanism design for large language models}. In \bibinfo{booktitle}{\emph{Proceedings of the ACM Web Conference 2024}}. \bibinfo{pages}{144--155}.
\newblock


\bibitem[D{\"u}tting et~al\mbox{.}(2019)]%
        {dutting2019optimal}
\bibfield{author}{\bibinfo{person}{Paul D{\"u}tting}, \bibinfo{person}{Zhe Feng}, \bibinfo{person}{Harikrishna Narasimhan}, \bibinfo{person}{David Parkes}, {and} \bibinfo{person}{Sai~Srivatsa Ravindranath}.} \bibinfo{year}{2019}\natexlab{}.
\newblock \showarticletitle{Optimal auctions through deep learning}. In \bibinfo{booktitle}{\emph{International Conference on Machine Learning}}. PMLR, \bibinfo{pages}{1706--1715}.
\newblock


\bibitem[Edelman et~al\mbox{.}(2007)]%
        {edelman2007internet}
\bibfield{author}{\bibinfo{person}{Benjamin Edelman}, \bibinfo{person}{Michael Ostrovsky}, {and} \bibinfo{person}{Michael Schwarz}.} \bibinfo{year}{2007}\natexlab{}.
\newblock \showarticletitle{Internet advertising and the generalized second-price auction: Selling billions of dollars worth of keywords}.
\newblock \bibinfo{journal}{\emph{American economic review}} \bibinfo{volume}{97}, \bibinfo{number}{1} (\bibinfo{year}{2007}), \bibinfo{pages}{242--259}.
\newblock


\bibitem[Gatti et~al\mbox{.}(2012)]%
        {gatti2012truthful}
\bibfield{author}{\bibinfo{person}{Nicola Gatti}, \bibinfo{person}{Alessandro Lazaric}, {and} \bibinfo{person}{Francesco Trovo}.} \bibinfo{year}{2012}\natexlab{}.
\newblock \showarticletitle{A truthful learning mechanism for contextual multi-slot sponsored search auctions with externalities}. In \bibinfo{booktitle}{\emph{Proceedings of the 13th ACM Conference on Electronic Commerce}}. \bibinfo{pages}{605--622}.
\newblock


\bibitem[Gloeckle et~al\mbox{.}(2024)]%
        {gloeckle2024better}
\bibfield{author}{\bibinfo{person}{Fabian Gloeckle}, \bibinfo{person}{Badr~Youbi Idrissi}, \bibinfo{person}{Baptiste Rozi{\`e}re}, \bibinfo{person}{David Lopez-Paz}, {and} \bibinfo{person}{Gabriel Synnaeve}.} \bibinfo{year}{2024}\natexlab{}.
\newblock \showarticletitle{Better \& faster large language models via multi-token prediction}.
\newblock \bibinfo{journal}{\emph{arXiv preprint arXiv:2404.19737}} (\bibinfo{year}{2024}).
\newblock


\bibitem[Gu and Sheng(2022)]%
        {gu2022ranking}
\bibfield{author}{\bibinfo{person}{Siyu Gu} {and} \bibinfo{person}{Xiangrong Sheng}.} \bibinfo{year}{2022}\natexlab{}.
\newblock \showarticletitle{On Ranking Consistency of Pre-ranking Stage}.
\newblock \bibinfo{journal}{\emph{arXiv preprint arXiv:2205.01289}} (\bibinfo{year}{2022}).
\newblock


\bibitem[Hummel and McAfee(2014)]%
        {hummel2014position}
\bibfield{author}{\bibinfo{person}{Patrick Hummel} {and} \bibinfo{person}{R~Preston McAfee}.} \bibinfo{year}{2014}\natexlab{}.
\newblock \showarticletitle{Position auctions with externalities}. In \bibinfo{booktitle}{\emph{Web and Internet Economics: 10th International Conference, WINE 2014, Beijing, China, December 14-17, 2014. Proceedings 10}}. Springer, \bibinfo{pages}{417--422}.
\newblock


\bibitem[Li et~al\mbox{.}(2024)]%
        {li2024deep}
\bibfield{author}{\bibinfo{person}{Xuejian Li}, \bibinfo{person}{Ze Wang}, \bibinfo{person}{Bingqi Zhu}, \bibinfo{person}{Fei He}, \bibinfo{person}{Yongkang Wang}, {and} \bibinfo{person}{Xingxing Wang}.} \bibinfo{year}{2024}\natexlab{}.
\newblock \showarticletitle{Deep automated mechanism design for integrating ad auction and allocation in feed}. In \bibinfo{booktitle}{\emph{Proceedings of the 47th International ACM SIGIR Conference on Research and Development in Information Retrieval}}. \bibinfo{pages}{1211--1220}.
\newblock


\bibitem[Liao et~al\mbox{.}(2022a)]%
        {liao2022nma}
\bibfield{author}{\bibinfo{person}{Guogang Liao}, \bibinfo{person}{Xuejian Li}, \bibinfo{person}{Ze Wang}, \bibinfo{person}{Fan Yang}, \bibinfo{person}{Muzhi Guan}, \bibinfo{person}{Bingqi Zhu}, \bibinfo{person}{Yongkang Wang}, \bibinfo{person}{Xingxing Wang}, {and} \bibinfo{person}{Dong Wang}.} \bibinfo{year}{2022}\natexlab{a}.
\newblock \showarticletitle{NMA: neural multi-slot auctions with externalities for online advertising}.
\newblock \bibinfo{journal}{\emph{arXiv preprint arXiv:2205.10018}} (\bibinfo{year}{2022}).
\newblock


\bibitem[Liao et~al\mbox{.}(2022b)]%
        {liao2022deep}
\bibfield{author}{\bibinfo{person}{Guogang Liao}, \bibinfo{person}{Xiaowen Shi}, \bibinfo{person}{Ze Wang}, \bibinfo{person}{Xiaoxu Wu}, \bibinfo{person}{Chuheng Zhang}, \bibinfo{person}{Yongkang Wang}, \bibinfo{person}{Xingxing Wang}, {and} \bibinfo{person}{Dong Wang}.} \bibinfo{year}{2022}\natexlab{b}.
\newblock \showarticletitle{Deep page-level interest network in reinforcement learning for ads allocation}. In \bibinfo{booktitle}{\emph{Proceedings of the 45th International ACM SIGIR Conference on Research and Development in Information Retrieval}}. \bibinfo{pages}{2292--2296}.
\newblock


\bibitem[Liao et~al\mbox{.}(2022c)]%
        {liao2022cross}
\bibfield{author}{\bibinfo{person}{Guogang Liao}, \bibinfo{person}{Ze Wang}, \bibinfo{person}{Xiaoxu Wu}, \bibinfo{person}{Xiaowen Shi}, \bibinfo{person}{Chuheng Zhang}, \bibinfo{person}{Yongkang Wang}, \bibinfo{person}{Xingxing Wang}, {and} \bibinfo{person}{Dong Wang}.} \bibinfo{year}{2022}\natexlab{c}.
\newblock \showarticletitle{Cross DQN: Cross deep Q network for ads allocation in feed}. In \bibinfo{booktitle}{\emph{Proceedings of the ACM Web Conference 2022}}. \bibinfo{pages}{401--409}.
\newblock


\bibitem[Liu et~al\mbox{.}(2024a)]%
        {liu2024deepseek}
\bibfield{author}{\bibinfo{person}{Aixin Liu}, \bibinfo{person}{Bei Feng}, \bibinfo{person}{Bing Xue}, \bibinfo{person}{Bingxuan Wang}, \bibinfo{person}{Bochao Wu}, \bibinfo{person}{Chengda Lu}, \bibinfo{person}{Chenggang Zhao}, \bibinfo{person}{Chengqi Deng}, \bibinfo{person}{Chenyu Zhang}, \bibinfo{person}{Chong Ruan}, {et~al\mbox{.}}} \bibinfo{year}{2024}\natexlab{a}.
\newblock \showarticletitle{Deepseek-v3 technical report}.
\newblock \bibinfo{journal}{\emph{arXiv preprint arXiv:2412.19437}} (\bibinfo{year}{2024}).
\newblock


\bibitem[Liu et~al\mbox{.}(2024b)]%
        {liu2024mmgrec}
\bibfield{author}{\bibinfo{person}{Han Liu}, \bibinfo{person}{Yinwei Wei}, \bibinfo{person}{Xuemeng Song}, \bibinfo{person}{Weili Guan}, \bibinfo{person}{Yuan-Fang Li}, {and} \bibinfo{person}{Liqiang Nie}.} \bibinfo{year}{2024}\natexlab{b}.
\newblock \showarticletitle{Mmgrec: Multimodal generative recommendation with transformer model}.
\newblock \bibinfo{journal}{\emph{arXiv preprint arXiv:2404.16555}} (\bibinfo{year}{2024}).
\newblock


\bibitem[Liu et~al\mbox{.}(2021)]%
        {liu2021neural}
\bibfield{author}{\bibinfo{person}{Xiangyu Liu}, \bibinfo{person}{Chuan Yu}, \bibinfo{person}{Zhilin Zhang}, \bibinfo{person}{Zhenzhe Zheng}, \bibinfo{person}{Yu Rong}, \bibinfo{person}{Hongtao Lv}, \bibinfo{person}{Da Huo}, \bibinfo{person}{Yiqing Wang}, \bibinfo{person}{Dagui Chen}, \bibinfo{person}{Jian Xu}, {et~al\mbox{.}}} \bibinfo{year}{2021}\natexlab{}.
\newblock \showarticletitle{Neural auction: End-to-end learning of auction mechanisms for e-commerce advertising}. In \bibinfo{booktitle}{\emph{Proceedings of the 27th ACM SIGKDD Conference on Knowledge Discovery \& Data Mining}}. \bibinfo{pages}{3354--3364}.
\newblock


\bibitem[Rajput et~al\mbox{.}(2023)]%
        {rajput2023recommender}
\bibfield{author}{\bibinfo{person}{Shashank Rajput}, \bibinfo{person}{Nikhil Mehta}, \bibinfo{person}{Anima Singh}, \bibinfo{person}{Raghunandan Hulikal~Keshavan}, \bibinfo{person}{Trung Vu}, \bibinfo{person}{Lukasz Heldt}, \bibinfo{person}{Lichan Hong}, \bibinfo{person}{Yi Tay}, \bibinfo{person}{Vinh Tran}, \bibinfo{person}{Jonah Samost}, {et~al\mbox{.}}} \bibinfo{year}{2023}\natexlab{}.
\newblock \showarticletitle{Recommender systems with generative retrieval}.
\newblock \bibinfo{journal}{\emph{Advances in Neural Information Processing Systems}}  \bibinfo{volume}{36} (\bibinfo{year}{2023}), \bibinfo{pages}{10299--10315}.
\newblock


\bibitem[Si et~al\mbox{.}(2024)]%
        {si2024generative}
\bibfield{author}{\bibinfo{person}{Zihua Si}, \bibinfo{person}{Zhongxiang Sun}, \bibinfo{person}{Jiale Chen}, \bibinfo{person}{Guozhang Chen}, \bibinfo{person}{Xiaoxue Zang}, \bibinfo{person}{Kai Zheng}, \bibinfo{person}{Yang Song}, \bibinfo{person}{Xiao Zhang}, \bibinfo{person}{Jun Xu}, {and} \bibinfo{person}{Kun Gai}.} \bibinfo{year}{2024}\natexlab{}.
\newblock \showarticletitle{Generative Retrieval with Semantic Tree-Structured Identifiers and Contrastive Learning}. In \bibinfo{booktitle}{\emph{Proceedings of the 2024 Annual International ACM SIGIR Conference on Research and Development in Information Retrieval in the Asia Pacific Region}}. \bibinfo{pages}{154--163}.
\newblock


\bibitem[Tan et~al\mbox{.}(2024)]%
        {tan2024idgenrec}
\bibfield{author}{\bibinfo{person}{Juntao Tan}, \bibinfo{person}{Shuyuan Xu}, \bibinfo{person}{Wenyue Hua}, \bibinfo{person}{Yingqiang Ge}, \bibinfo{person}{Zelong Li}, {and} \bibinfo{person}{Yongfeng Zhang}.} \bibinfo{year}{2024}\natexlab{}.
\newblock \showarticletitle{Idgenrec: Llm-recsys alignment with textual id learning}. In \bibinfo{booktitle}{\emph{Proceedings of the 47th International ACM SIGIR Conference on Research and Development in Information Retrieval}}. \bibinfo{pages}{355--364}.
\newblock


\bibitem[Tang et~al\mbox{.}(2023)]%
        {tang2023recent}
\bibfield{author}{\bibinfo{person}{Yubao Tang}, \bibinfo{person}{Ruqing Zhang}, \bibinfo{person}{Jiafeng Guo}, {and} \bibinfo{person}{Maarten de Rijke}.} \bibinfo{year}{2023}\natexlab{}.
\newblock \showarticletitle{Recent advances in generative information retrieval}. In \bibinfo{booktitle}{\emph{Proceedings of the Annual International ACM SIGIR Conference on Research and Development in Information Retrieval in the Asia Pacific Region}}. \bibinfo{pages}{294--297}.
\newblock


\bibitem[Tay et~al\mbox{.}(2022)]%
        {tay2022transformer}
\bibfield{author}{\bibinfo{person}{Yi Tay}, \bibinfo{person}{Vinh Tran}, \bibinfo{person}{Mostafa Dehghani}, \bibinfo{person}{Jianmo Ni}, \bibinfo{person}{Dara Bahri}, \bibinfo{person}{Harsh Mehta}, \bibinfo{person}{Zhen Qin}, \bibinfo{person}{Kai Hui}, \bibinfo{person}{Zhe Zhao}, \bibinfo{person}{Jai Gupta}, {et~al\mbox{.}}} \bibinfo{year}{2022}\natexlab{}.
\newblock \showarticletitle{Transformer memory as a differentiable search index}.
\newblock \bibinfo{journal}{\emph{Advances in Neural Information Processing Systems}}  \bibinfo{volume}{35} (\bibinfo{year}{2022}), \bibinfo{pages}{21831--21843}.
\newblock


\bibitem[Wang et~al\mbox{.}(2022b)]%
        {wang2022designing}
\bibfield{author}{\bibinfo{person}{Yiqing Wang}, \bibinfo{person}{Xiangyu Liu}, \bibinfo{person}{Zhenzhe Zheng}, \bibinfo{person}{Zhilin Zhang}, \bibinfo{person}{Miao Xu}, \bibinfo{person}{Chuan Yu}, {and} \bibinfo{person}{Fan Wu}.} \bibinfo{year}{2022}\natexlab{b}.
\newblock \showarticletitle{On designing a two-stage auction for online advertising}. In \bibinfo{booktitle}{\emph{Proceedings of the ACM Web Conference 2022}}. \bibinfo{pages}{90--99}.
\newblock


\bibitem[Wang et~al\mbox{.}(2024)]%
        {wang2024content}
\bibfield{author}{\bibinfo{person}{Yidan Wang}, \bibinfo{person}{Zhaochun Ren}, \bibinfo{person}{Weiwei Sun}, \bibinfo{person}{Jiyuan Yang}, \bibinfo{person}{Zhixiang Liang}, \bibinfo{person}{Xin Chen}, \bibinfo{person}{Ruobing Xie}, \bibinfo{person}{Su Yan}, \bibinfo{person}{Xu Zhang}, \bibinfo{person}{Pengjie Ren}, {et~al\mbox{.}}} \bibinfo{year}{2024}\natexlab{}.
\newblock \showarticletitle{Content-Based Collaborative Generation for Recommender Systems}. In \bibinfo{booktitle}{\emph{Proceedings of the 33rd ACM International Conference on Information and Knowledge Management}}. \bibinfo{pages}{2420--2430}.
\newblock


\bibitem[Wang et~al\mbox{.}(2022a)]%
        {wang2022learning}
\bibfield{author}{\bibinfo{person}{Ze Wang}, \bibinfo{person}{Guogang Liao}, \bibinfo{person}{Xiaowen Shi}, \bibinfo{person}{Xiaoxu Wu}, \bibinfo{person}{Chuheng Zhang}, \bibinfo{person}{Yongkang Wang}, \bibinfo{person}{Xingxing Wang}, {and} \bibinfo{person}{Dong Wang}.} \bibinfo{year}{2022}\natexlab{a}.
\newblock \showarticletitle{Learning List-wise Representation in Reinforcement Learning for Ads Allocation with Multiple Auxiliary Tasks}. In \bibinfo{booktitle}{\emph{Proceedings of the 31st ACM International Conference on Information \& Knowledge Management}}. \bibinfo{pages}{3555--3564}.
\newblock


\bibitem[Xie et~al\mbox{.}(2021)]%
        {xie2021hierarchical}
\bibfield{author}{\bibinfo{person}{Ruobing Xie}, \bibinfo{person}{Shaoliang Zhang}, \bibinfo{person}{Rui Wang}, \bibinfo{person}{Feng Xia}, {and} \bibinfo{person}{Leyu Lin}.} \bibinfo{year}{2021}\natexlab{}.
\newblock \showarticletitle{Hierarchical reinforcement learning for integrated recommendation}. In \bibinfo{booktitle}{\emph{Proceedings of the AAAI conference on artificial intelligence}}, Vol.~\bibinfo{volume}{35}. \bibinfo{pages}{4521--4528}.
\newblock


\bibitem[Xu et~al\mbox{.}(2023)]%
        {xu2023multi}
\bibfield{author}{\bibinfo{person}{Yue Xu}, \bibinfo{person}{Qijie Shen}, \bibinfo{person}{Jianwen Yin}, \bibinfo{person}{Zengde Deng}, \bibinfo{person}{Dimin Wang}, \bibinfo{person}{Hao Chen}, \bibinfo{person}{Lixiang Lai}, \bibinfo{person}{Tao Zhuang}, {and} \bibinfo{person}{Junfeng Ge}.} \bibinfo{year}{2023}\natexlab{}.
\newblock \showarticletitle{Multi-channel Integrated Recommendation with Exposure Constraints}. In \bibinfo{booktitle}{\emph{Proceedings of the 29th ACM SIGKDD Conference on Knowledge Discovery and Data Mining}}. \bibinfo{pages}{5338--5349}.
\newblock


\bibitem[Yan et~al\mbox{.}(2020)]%
        {yan2020ads}
\bibfield{author}{\bibinfo{person}{Jinyun Yan}, \bibinfo{person}{Zhiyuan Xu}, \bibinfo{person}{Birjodh Tiwana}, {and} \bibinfo{person}{Shaunak Chatterjee}.} \bibinfo{year}{2020}\natexlab{}.
\newblock \showarticletitle{Ads allocation in feed via constrained optimization}. In \bibinfo{booktitle}{\emph{Proceedings of the 26th ACM SIGKDD International Conference on Knowledge Discovery \& Data Mining}}. \bibinfo{pages}{3386--3394}.
\newblock


\bibitem[Yang et~al\mbox{.}(2025)]%
        {yang2025sparse}
\bibfield{author}{\bibinfo{person}{Yuhao Yang}, \bibinfo{person}{Zhi Ji}, \bibinfo{person}{Zhaopeng Li}, \bibinfo{person}{Yi Li}, \bibinfo{person}{Zhonglin Mo}, \bibinfo{person}{Yue Ding}, \bibinfo{person}{Kai Chen}, \bibinfo{person}{Zijian Zhang}, \bibinfo{person}{Jie Li}, \bibinfo{person}{Shuanglong Li}, {et~al\mbox{.}}} \bibinfo{year}{2025}\natexlab{}.
\newblock \showarticletitle{Sparse Meets Dense: Unified Generative Recommendations with Cascaded Sparse-Dense Representations}.
\newblock \bibinfo{journal}{\emph{arXiv preprint arXiv:2503.02453}} (\bibinfo{year}{2025}).
\newblock


\bibitem[Yang et~al\mbox{.}(2024)]%
        {yang2024parallel}
\bibfield{author}{\bibinfo{person}{Zhiguang Yang}, \bibinfo{person}{Liufang Sang}, \bibinfo{person}{Haoran Wang}, \bibinfo{person}{Wenlong Chen}, \bibinfo{person}{Lu Wang}, \bibinfo{person}{Jie He}, \bibinfo{person}{Changping Peng}, \bibinfo{person}{Zhangang Lin}, \bibinfo{person}{Chun Gan}, {and} \bibinfo{person}{Jingping Shao}.} \bibinfo{year}{2024}\natexlab{}.
\newblock \showarticletitle{Parallel ranking of ads and creatives in real-time advertising systems}. In \bibinfo{booktitle}{\emph{Proceedings of the AAAI Conference on Artificial Intelligence}}, Vol.~\bibinfo{volume}{38}. \bibinfo{pages}{9278--9286}.
\newblock


\bibitem[Zeghidour et~al\mbox{.}(2021)]%
        {zeghidour2021soundstream}
\bibfield{author}{\bibinfo{person}{Neil Zeghidour}, \bibinfo{person}{Alejandro Luebs}, \bibinfo{person}{Ahmed Omran}, \bibinfo{person}{Jan Skoglund}, {and} \bibinfo{person}{Marco Tagliasacchi}.} \bibinfo{year}{2021}\natexlab{}.
\newblock \showarticletitle{Soundstream: An end-to-end neural audio codec}.
\newblock \bibinfo{journal}{\emph{IEEE/ACM Transactions on Audio, Speech, and Language Processing}}  \bibinfo{volume}{30} (\bibinfo{year}{2021}), \bibinfo{pages}{495--507}.
\newblock


\bibitem[Zhai et~al\mbox{.}(2024)]%
        {zhai2024actions}
\bibfield{author}{\bibinfo{person}{Jiaqi Zhai}, \bibinfo{person}{Lucy Liao}, \bibinfo{person}{Xing Liu}, \bibinfo{person}{Yueming Wang}, \bibinfo{person}{Rui Li}, \bibinfo{person}{Xuan Cao}, \bibinfo{person}{Leon Gao}, \bibinfo{person}{Zhaojie Gong}, \bibinfo{person}{Fangda Gu}, \bibinfo{person}{Michael He}, {et~al\mbox{.}}} \bibinfo{year}{2024}\natexlab{}.
\newblock \showarticletitle{Actions speak louder than words: Trillion-parameter sequential transducers for generative recommendations}.
\newblock \bibinfo{journal}{\emph{arXiv preprint arXiv:2402.17152}} (\bibinfo{year}{2024}).
\newblock


\bibitem[Zhang(2021)]%
        {zhang2021survey}
\bibfield{author}{\bibinfo{person}{Zhanhao Zhang}.} \bibinfo{year}{2021}\natexlab{}.
\newblock \showarticletitle{A survey of online auction mechanism design using deep learning approaches}.
\newblock \bibinfo{journal}{\emph{arXiv preprint arXiv:2110.06880}} (\bibinfo{year}{2021}).
\newblock


\bibitem[Zhang et~al\mbox{.}(2023)]%
        {zhang2023rethinking}
\bibfield{author}{\bibinfo{person}{Zhixuan Zhang}, \bibinfo{person}{Yuheng Huang}, \bibinfo{person}{Dan Ou}, \bibinfo{person}{Sen Li}, \bibinfo{person}{Longbin Li}, \bibinfo{person}{Qingwen Liu}, {and} \bibinfo{person}{Xiaoyi Zeng}.} \bibinfo{year}{2023}\natexlab{}.
\newblock \showarticletitle{Rethinking the role of pre-ranking in large-scale e-commerce searching system}.
\newblock \bibinfo{journal}{\emph{arXiv preprint arXiv:2305.13647}} (\bibinfo{year}{2023}).
\newblock


\bibitem[Zhang et~al\mbox{.}(2021)]%
        {zhang2021optimizing}
\bibfield{author}{\bibinfo{person}{Zhilin Zhang}, \bibinfo{person}{Xiangyu Liu}, \bibinfo{person}{Zhenzhe Zheng}, \bibinfo{person}{Chenrui Zhang}, \bibinfo{person}{Miao Xu}, \bibinfo{person}{Junwei Pan}, \bibinfo{person}{Chuan Yu}, \bibinfo{person}{Fan Wu}, \bibinfo{person}{Jian Xu}, {and} \bibinfo{person}{Kun Gai}.} \bibinfo{year}{2021}\natexlab{}.
\newblock \showarticletitle{Optimizing multiple performance metrics with deep GSP auctions for e-commerce advertising}. In \bibinfo{booktitle}{\emph{Proceedings of the 14th ACM International Conference on Web Search and Data Mining}}. \bibinfo{pages}{993--1001}.
\newblock


\bibitem[Zhao et~al\mbox{.}(2021)]%
        {zhao2021dear}
\bibfield{author}{\bibinfo{person}{Xiangyu Zhao}, \bibinfo{person}{Changsheng Gu}, \bibinfo{person}{Haoshenglun Zhang}, \bibinfo{person}{Xiwang Yang}, \bibinfo{person}{Xiaobing Liu}, \bibinfo{person}{Jiliang Tang}, {and} \bibinfo{person}{Hui Liu}.} \bibinfo{year}{2021}\natexlab{}.
\newblock \showarticletitle{Dear: Deep reinforcement learning for online advertising impression in recommender systems}. In \bibinfo{booktitle}{\emph{Proceedings of the AAAI conference on artificial intelligence}}, Vol.~\bibinfo{volume}{35}. \bibinfo{pages}{750--758}.
\newblock


\bibitem[Zhao et~al\mbox{.}(2023)]%
        {zhao2023copr}
\bibfield{author}{\bibinfo{person}{Zhishan Zhao}, \bibinfo{person}{Jingyue Gao}, \bibinfo{person}{Yu Zhang}, \bibinfo{person}{Shuguang Han}, \bibinfo{person}{Siyuan Lou}, \bibinfo{person}{Xiang-Rong Sheng}, \bibinfo{person}{Zhe Wang}, \bibinfo{person}{Han Zhu}, \bibinfo{person}{Yuning Jiang}, \bibinfo{person}{Jian Xu}, {et~al\mbox{.}}} \bibinfo{year}{2023}\natexlab{}.
\newblock \showarticletitle{COPR: Consistency-Oriented Pre-Ranking for Online Advertising}. In \bibinfo{booktitle}{\emph{Proceedings of the 32nd ACM International Conference on Information and Knowledge Management}}. \bibinfo{pages}{4974--4980}.
\newblock


\bibitem[Zheng et~al\mbox{.}(2024)]%
        {zheng2024adapting}
\bibfield{author}{\bibinfo{person}{Bowen Zheng}, \bibinfo{person}{Yupeng Hou}, \bibinfo{person}{Hongyu Lu}, \bibinfo{person}{Yu Chen}, \bibinfo{person}{Wayne~Xin Zhao}, \bibinfo{person}{Ming Chen}, {and} \bibinfo{person}{Ji-Rong Wen}.} \bibinfo{year}{2024}\natexlab{}.
\newblock \showarticletitle{Adapting large language models by integrating collaborative semantics for recommendation}. In \bibinfo{booktitle}{\emph{2024 IEEE 40th International Conference on Data Engineering (ICDE)}}. IEEE, \bibinfo{pages}{1435--1448}.
\newblock


\bibitem[Zhu et~al\mbox{.}(2024)]%
        {zhu2024contextual}
\bibfield{author}{\bibinfo{person}{Ruitao Zhu}, \bibinfo{person}{Yangsu Liu}, \bibinfo{person}{Dagui Chen}, \bibinfo{person}{Zhenjia Ma}, \bibinfo{person}{Chufeng Shi}, \bibinfo{person}{Zhenzhe Zheng}, \bibinfo{person}{Jie Zhang}, \bibinfo{person}{Jian Xu}, \bibinfo{person}{Bo Zheng}, {and} \bibinfo{person}{Fan Wu}.} \bibinfo{year}{2024}\natexlab{}.
\newblock \showarticletitle{Contextual Generative Auction with Permutation-level Externalities for Online Advertising}.
\newblock \bibinfo{journal}{\emph{arXiv preprint arXiv:2412.11544}} (\bibinfo{year}{2024}).
\newblock


\end{thebibliography}

\end{document}